# Robust Optical Data Encryption by Projection-Photoaligned Polymer-Stabilized-Liquid-Crystals

*Siying Liu[1,2], Saleh Alfarhan[2], Wenbo Wang[1], Shuai Feng[2], Yuxiang Zhu[1], Luyang Liu[1], Kenan Song[1], Sui Yang[2]*, **Kailong Jin[2]\***, *Xiangfan Chen[1]\**

1. School of Manufacturing Systems and Networks, Arizona State University, Mesa, AZ 85212, US.
2. School for Engineering of Matter, Transport & Energy, Arizona State University, Tempe, AZ 85287, US

Corresponding Authors: Kailong Jin: Kailong.Jin@asu.edu; Xiangfan Chen: Xiangfan.Chen@asu.edu

**Abstract:** The emerging Internet of Things (IoTs) invokes increasing security demands that require robust encryption or anti-counterfeiting technologies. Albeit being acknowledged as efficacious solutions in processing elaborate graphical information via multiple degrees of freedom, optical data encryption and anti-counterfeiting techniques are typically inept in delivering satisfactory performance without compromising the desired ease-of-processibility or compatibility, thus leading to the exploration of novel materials and devices that are competent. Here, a robust optical data encryption technique is demonstrated utilizing polymer-stabilized-liquid-crystals (PSLCs) combined with projection photoalignment and photopatterning methods. The PSLCs possess implicit optical patterns encoded via photoalignment, as well as explicit geometries produced via photopatterning. Furthermore, the PSLCs demonstrate improved robustness against harsh chemical environments and thermal stability, and can be directly deployed onto various rigid and flexible substrates. Based on this, it is demonstrated that single PSLC is apt to carry intricate information, or serve as exclusive watermark with both implicit features and explicit geometries. Moreover, a novel, generalized design strategy is developed, for the first time, to encode intricate and exclusive information with enhanced security by spatially programming the photoalignment patterns of a pair of cascade PSLCs, which further illustrates the promising capabilities of PSLCs in optical data encryption and anti-counterfeiting.

# 1. Introduction

The prosperous thriving of the Internet of Things (IoTs) puts forth critical requirements on information security and anti-counterfeiting systems.[1,2] Among the vast categories of prevailing techniques, optical encryption is prominent for applications such as authentication or identification of devices, currencies or valuable commodities, data storage, dissemination or other cryptographic purposes.[3,4] For instance, optical metasurfaces that comprise ultrathin optical elements capable of manipulating the amplitudes, phases, polarizations or orbital angular momentums of the lightwaves based on plasmonic resonances, Pancharatnam-Berry (PB) phases, Mie resonances, *etc.* have been explored as potential solutions for information security. Particularly, static or dynamic graphical information can be encoded onto a plethora of optical metasurfaces via independent or combined manipulations of ligh at single or multi-wavelengths.[5–17] Nevertheless, they are often renowned for their intricate designs that lack versatility, tedious fabrication and associated complex characterization apparatus, which inevitably jeopardize their potential in broader practical applications.[18–23] Other optical encryption methodologies such as physically unclonable functions (PUFs)[24–26] and optical double random phase encoding (DRPE)[27–29] often suffer from similar dilemma.

On the other hand, liquid crystals (LCs) possessing optical and dielectric anisotropy generated by facile processing routes provide an ideal platform for optical encryption devices,[30] and have been extensively used in the fields such as imaging and displays,[31,32] optical anti-counterfeiting,[33–35] holography,[36] augmented and virtual reality.[37] Particularly, in the polymer-stabilized-liquid-crystals (PSLCs), the low-molar-mass LC molecules serve as the anisotropic matrix, and their alignment is stabilized via their elastic interaction with the embedded and polymerized mesogenic monomers.[38–40] Functional devices fabricated from PSLCs usually require spatially programmable orientational directors, which are attainable via alignment techniques involving the assistance of mechanical stress,[41] electrical field,[42] magnetic field,[43] geometric confinement,[44,45] or polarized illuminations, and the last category is referred to as photoalignment.[46] Especially, photoalignment is typically conducted by either direct doping of optically anisotropic chromophores such as azobenzene-based dye molecules into the LC matrices, or via substrates modified by dye molecules, followed by the irradiation of linearly polarized light. The photoswitchable azobenzene groups tend to reorient perpendicularly with regard to the polarizations of light via repeated *trans-cis-trans* isomerization, and such alignment is subsequently transferred to the LCs via intermolecular interaction.[46–49] Compared to other alignment methods, photoalignment is favorable due to its satisfactory reversibility and compatibility with delicate optical systems to produce pixel-level, programmable alignment patterns,[50,51] which open up new opportunities for optical encryption and anti-counterfeiting devices. For instance, He *et al.* developed chemically synthesized, dual-mode LC polymers that possessed both polarization and photoluminescent patterns via photoalignment through masked UV irradiation.[52] Sang *et al.* demonstrated that photoaligned PSLCs can be readily prepared by prescribed photomasks.[53] Liu *et al.* fabricated polymer-stabilized heliconical structures and achieved spectral tunings via external electric fields.[54] However, those works still heavily relied on prescribed photomasks, laborious chemical syntheses, or sophisticated characterization apparatus. More importantly, information security in these works is still notably vulnerable to theft and can be decrypted in a relatively easy manner, such as a rotational polarizer, because all the essential elements constituting the encoded information are carried by a single device. Further boosting of the information security by distributing the essential elements to encoders and paired decoding devices / beam profiles would undesirably require dynamic pattern designs and device fabrications that are much more complex and effort-consuming, thus further limiting the practical applications. Therefore, the appealing potential of photoaligned PSLCs in information encryption or storage remains to be further exploited.

Herein, as illustrated in **Figure 1**, we demonstrate a new robust optical data encryption and anti-counterfeiting strategy enabled by photoaligned PSLCs. Those PSLCs-based devices can be achieved by combining projection-enabled photoalignment and photopatterning of LCs with dynamic masks (**Figure 1a & Figure S1**). As schemed in **Figure 1a**, the azobenzene dye, Brilliant Yellow (BY), can be firstly aligned along desired directions by polarized optical patterns through repeated *trans-cis-trans* isomerization (**i**), which can then serve as "commanding" layer to align the directors ($\vec{n}$) of the LC molecules via intermolecular interaction (**ii**). The photocurable LC mixtures comprising non-reactive and reactive LC mesogens can be readily photopolymerized into sophistically pattened PSLCs (**iii**) to maintain the aligned directors ($\vec{n}$). Their photoaligned patterns can be retained even after the embedded non-reactive LC molecules are extracted. Therefore, these photopatterned PSLCs not only possess explicit, solid geometries, but also implicit, polarized optical patterns (**Figure S2a**) and can be vigorously bonded onto either rigid or flexible substrates (**iii**).

With the developed procedures, we demonstrate several paradigms that unveil the potential of the PSLCs-based devices in information storage and dissemination, as well as anti-counterfeiting. We first show that single PSLC is potent to serve as anti-counterfeiting watermarks (**Figure 1a(ii)**) or carry elaborate information such as QR code (**Figure 1b(i)** and **Figure S2b**) which can only be identified by the Polarized Optical Microscopy (POM). Most importantly, for the first time, a novel strategy to yield spatially programmable optical data encryption via a pair of complementary, cascade PSLCs is developed (**Figure 1b(ii)**). Via paired PSLCs, information security can be significantly enhanced because the elements necessarily required to disclose the encrypted information are distributed to matched pairs, hence making them challenging to counterfeit and much less vulnerable against theft or loss. Down this path, we first demonstrate a special case and validate that we can spatially program the cascade PSLCs' output to be "on" (annotated as "*1*") or "off" (annotated as "*0*", all the angles here-and-after were defined regarding the first linear polarizer in the characterization setup) by stacking the PSLCs with directors ($\vec{n}$) along *45°* (annotated as "*+1*") and *135°* (annotated as "*-1*") which are visually identical in POM. As a result, we can implement desired information manipulation (e.g., hide, show, and mislead). Moreover, we analytically and experimentally confirm that generalized spectral tunings are readily achievable upon the cascade PSLCs with various combinations of directors ($\vec{n}$), which further broaden the versatile application of PSLCs in optical data encryption and anti-counterfeiting.

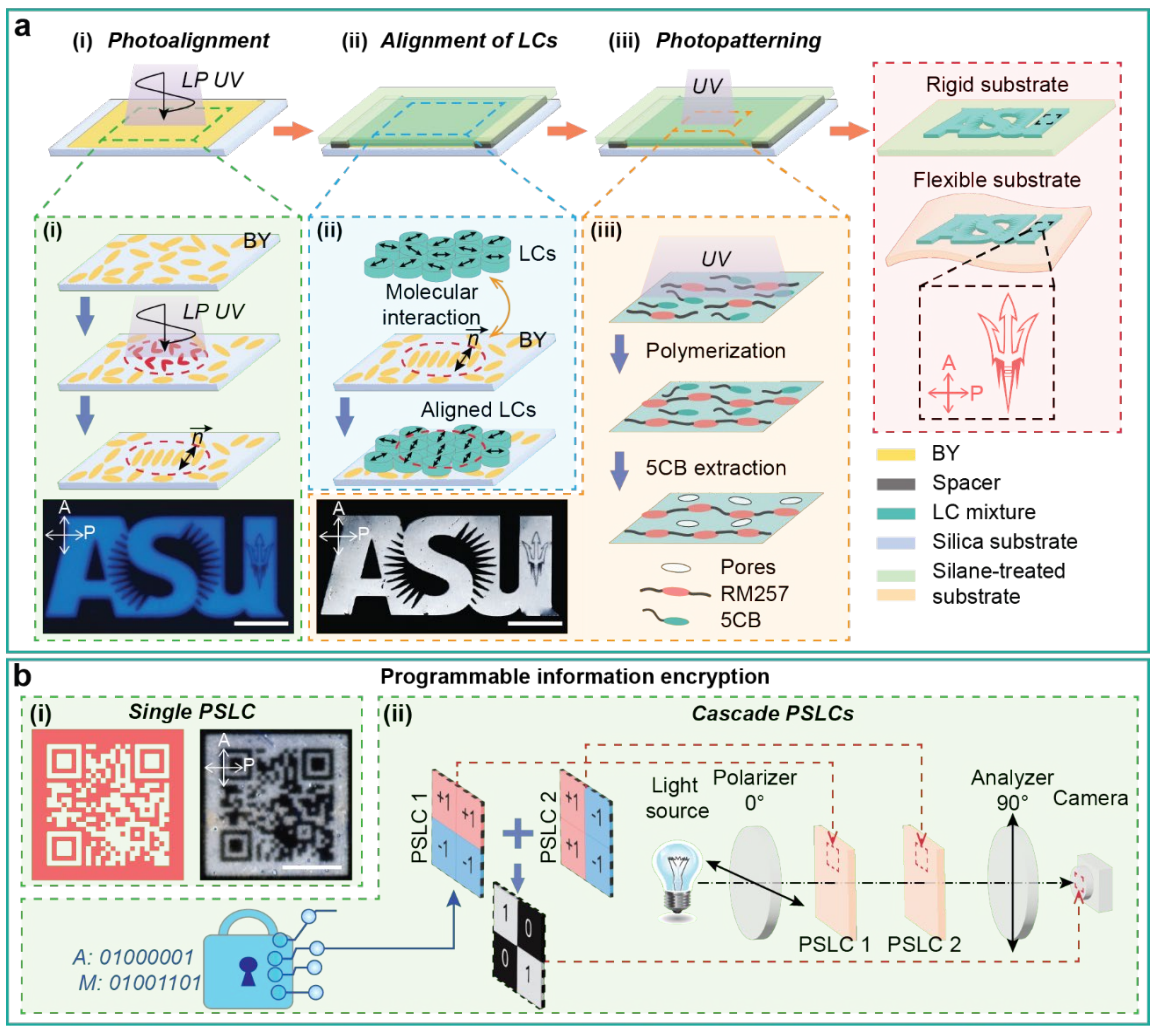

**Figure 1. Schematic illustration of projection photoalignment and photopatterning of PSLCs for optical anti-counterfeiting and data encryption.** (a) Fabrication flow, associated mechanisms, and corresponding results of PSLCs bonded onto different substrates, including: (i) Photoalignment of the azobenzene dye, i.e., Brilliant Yellow (BY), inset: Polarized Optical Microscopy (POM) image of photoaligned BY; (ii) Transfer of the aligned patterns from BY to LC molecules via intermolecular interaction; (iii) Photopatterning of aligned LCs, and subsequent extraction of the embedded non-reactive LC molecules, inset: POM image of the photoaligned and photopatterned PSLC. (b) Schematic of programmable information encryption via PSLCs, including (i) single PSLC chip only identifiable under POM, and (ii) complementary, cascade PSLCs. Inset in (i): projection pattern (left) and POM image (right) of the example QR code encoded via photoalignment. Scale bars: 2 mm.

## 2. Results

### 2.1. Photoalignment and Photopatterning of PSLCs

To yield photoalignment in a pixelated manner, a customized optical setup based on a 405 nm UV projector comprising a Digital Mirror Device (DMD) to generate dynamic masks (**Figure S1** and **Experimental Section**) was built.[55,56] A UV linear polarizer mounted atop a motorized rotary stage was included to control the polarizations of the projected patterns, hence enabled localized photoalignment with arbitrary director ($\vec{n}$). Accordingly, BY (chemical structure in **Figure S3**) was selected as the azobenzene-based photoalignment dye because of its appropriate working wavelength as well as promising anchoring energy.[57,58] The photoalignment substrates were prepared by spin-coating of BY solution onto oxygen-plasma treated substrates, followed by linearly polarized and patterned UV irradiations (**Experimental Section**). A series of photoalignment substrates with complex patterns (**Figure S2c-e**), such as the "pitchfork" (**Figure 2a$_1$**), were prepared and examined using POM (**Experimental Section**). Due to the angle-dependent absorbing characteristics,[59] the photoaligned BY in POM images appear as patterns with varying brightness, which can be well indicated by the concentric rings aligned with different angles from *0°* to (the brightest ring) to *45°* (the darkest ring), (**Figure 2a$_2$** and **Figure S2d**). Moreover, benefiting from the high-resolution optical setup (lateral resolution of 8.9 × 8.9 μm$^2$ pixel$^{-1}$), it is also facile to prepare photoalignment substrates comprising aligned features down to micrometer-scale (smallest feature ~ 50 μm), as shown in hierarchical chessboard patterns (**Figure 2a$_3$,a$_4$**),.

To facilitate the transfer of the photoaligned patterns from BY to the LCs and obtain fine solidified LC features, low-molar-mass, non-reactive LC, 4-Cyano-4'-pentylbiphenyl (5CB), and reactive mesogenic monomer, 2-Methyl-1,4-phenylene bis(4-(3-(acryloyloxy)propoxy)benzoate) (RM257, chemical structures of 5CB and RM257 in **Figure**

S3), as well as photo-initiator and photo-absorber, were utilized to formulate photocurable LC mixtures. The LC mixtures manifest relatively low nematic-to-isotropic temperature $T_{NI}$ ($T_{NI} = 54.2\ °C$, determined by Differential Scanning Calorimetry, DSC) and low viscosities both at nematic and isotropic states (**Experimental Section** and **Figure S4**). As schemed in **Figure 1a**, photoaligned and photopatterned PSLCs bonded onto various substrates were achieved by constructing two-dimensional (2D) cells comprising of one photoalignment substrate and one silane-treated substrate,[55] with precisely controlled cell gaps (5 μm, **Experimental Section**). As-prepared photocurable LC mixtures were injected into the cells via capillary action at elevated temperature well above the $T_{NI}$ and were then aligned by the photoalignment substrates after cooling down to room temperature. Finally, the LC mixtures were cured by patterned UV illuminations using the same optical setup (**Experimental Section**). Then, we extracted the embedded non-reactive 5CB using organic solvents (i.e., N,N-Dimethylformamide, and ethanol), leaving only the polymerized RM257 networks behind (**Experimental Section**). Upon optimized photocurable LC mixtures, we were capable of patterning well-defined, complex geometries, such as "walking cat", whose optical transparency can be both visually and quantitatively characterized using UV-Vis spectrometry, while its morphology can by verified by optical profilometer (**Experimental Section** and **Figure S5**). The PSLCs possess superior transparency (>90%) within the visible regime under unpolarized illumination, which ensures the inclusion of the PSLCs into devices that demand overall transparency to retain designated anti-counterfeiting purposes. Overall, via the customized setup and optimized procedure, photopatterned PSLCs retained the complex patterns from the photoaligned BY via the intermolecular interaction between BY and LC molecules, as implied by the POM images of PLSCs (**Figure 2b**) prepared using the photoalignment substrates corresponding to those in **Figure 2a**. It is worth noting that the micron-scale photoaligned pattern was also successfully transferred to the PSLC, as depicted in **Figure 2b$_4$**, which further grants the potential of the developed procedures in the multi-scale fabrication of photoaligned LC devices for diverse applications..

To quantitatively study the anisotropic optical properties of the aligned PSLCs, the polarized transmission spectra of unidirectionally photoaligned and photopatterned PSLC film were measured throughout the entire polar quadrants with a *5°*-interval (**Experimental Section** and **Figure S6**). A representative set of complete spectra from *0°* to *45°* was shown in **Figure 2c**, and the corresponding transmission at 532 nm was polar plotted in **Figure 2d**. These spectral characterizations explicitly reveal the angle-dependent transmitting characteristics of LCs, with maxima at *45°, 135°, 225°* and *315°,* and minima at *0°, 90°, 180°* and *270°,* respectively,[60] confirming the optical anisotropy induced by well-aligned orientations.

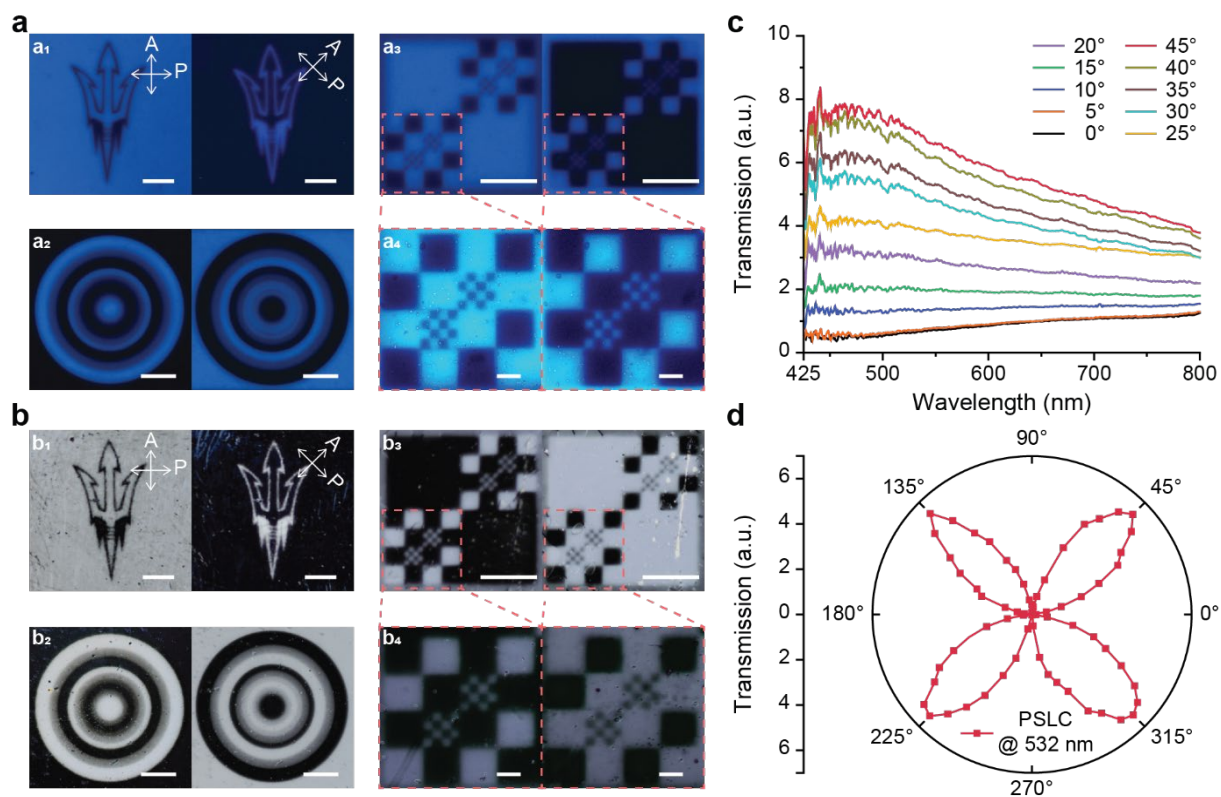

**Figure 2. Photoaligned and photopatterned PSLCs.** (a) POM images of the photoaligned BY, including ASU "Pitchfork" (a1), concentric rings (a2), hierarchical chessboard (a3, a4). Scale bars: 1 mm (a1-a3) and 200 μm (a4). (b) POM images of the photoaligned and photopatterned PSLCs comprising patterns in (a). Scale bars: 1 mm (b1-b3) and 200 μm (b4). (c) Polarized transmission spectra of unidirectionally photoaligned and photopatterned PSLC from *0°* to *45°*. (d) Polar-plotted polarized transmission spectra at 532 nm.

## 2.2. Enhanced Environmental Robustness and Broadened Range-of-Applicability of PSLCs

To better understand the compositional and structural properties of the PSLCs and unveil the origin of the complex photoalignment patterns after 5CB extraction, we first adopted Fourier-Transform Infrared Spectroscopy (FTIR) to verify the removal of 5CB. As implied in **Figure 3a**, both peaks at 1732 cm$^{-1}$ ($C = O$ stretching, solely on RM257) and 2227 cm$^{-1}$ ($C \equiv N$ stretching, solely on 5CB) were present before 5CB extraction, whereas the latter peak disappeared after the 5CB extraction, indicating the complete removal of 5CB.[61] Additionally, Scanning Electron Microscopy (SEM) was employed to characterize the surface morphologies of PSLCs encoded with complex pattern, such as the "cat family", whose directors ($\vec{n}$) were aligned along different angles (*0°, 22.5°, 45°*) at designated regions (**Figure S2f** and **Figure**

**3b**). SEM images of corresponding surface regions on the PSLC (**Figure 3b**) evidently reveal the pores approximately along *0°, 22.5°* and *45°*, respectively. These pores originated from the extraction of the 5CB embedded within the photoaligned and polymerized RM257 networks and implies the successful transfer of the photoalignment patterns from BY to the LCs. After photopatterning, the photoalignment patterns were permanently fixed due to the polymerized RM257 network.[62] Because of the crosslinked network, as-fabricated PSLCs after 5CB extraction (i.e., neat crosslinked RM257) manifest excellent tolerance against harsh organic/aqueous environment as well as thermal stability without sacrificing their designed functionality. As indicated by the DSC thermogram (**Figure 3c**), the crosslinked RM257 network showed no obvious phase transitions when heated up to 200 °C, consistent with previous studies.[62] The PSLCs only started to decompose at a much higher temperature of ~345 °C, as confirmed by Thermal Gravimetric Analysis (TGA, **Experimental Section** and **Figure S7**). Thermal stability test on the 5CB-extracted PSLCs (**Experimental Section**, **Figure 3d,e** and **Figure S8**) also validate that the photoaligned patterns were well preserved even after heating at ~200 °C. In addition, to test the chemical stability of the PSLCs under harsh conditions, the same sample was subject to a variety of non-polar/polar, organic/inorganic solvents or pH buffers (i.e., ethanol, dichloromethane, toluene, acidic and basic buffers with pH at 3.00 and 11.00, respectively), and its intactness was confirmed by the POM images (**Experimental Section** and **Figure S9**). Owing to the superior chemical stability of the polymerized RM257 networks as well as the robust interfacial bonding between the PSLCs and the silane-treated substrates,[63,64] the PSLC was able to preserve its initial photoalignment pattern without any noticeable degradation or unwanted detachment from the substrate that could impair its designed functionality (**Figure 3e**). These tests indicate the appealing reliability of the PSLCs in serving optical encryption purposes under harsh environmental conditions.

On the other side, to further broaden the range-of-applicability of the photoaligned and photopatterned PSLCs, we verified the ease-of-deployment of the PSLCs by directly fabricating them on various silane-treated substrates, including rigid substrates such as silica and c-plane sapphire, or flexible substrates such as bare polyethylene terephthalate (PET) or indium-tin-oxide(ITO)-coated PET films (**Experimental Section**).[65,66] With this facile silanization treatment, we effectively eliminated the necessity of any post transferring procedures that calls for extra work load and undermines the device delicacy. POM images of the PSLCs bonded on different substrates (**Figure 3f**) show that these PSLCs well preserved their photoalignment patterns (i.e., "sun devil" logo), except the defects and colorful wrinkles that could be attributed to the less uniform cell assembly using flexible substrates as well as the intrinsic chain alignments within the as-received PET and ITO-coated PET films (**Figure S10**). Especially, the PLSCs bonded onto flexible substrates can even retain their photoalignment patterns at highly bent state (**Figure 3g** and **Figure S11**). Hence, the direct integration of PSLC with flexible, conductive substrates further grants the possibilities in designing novel optical devices that are flexible and electrically tunable.

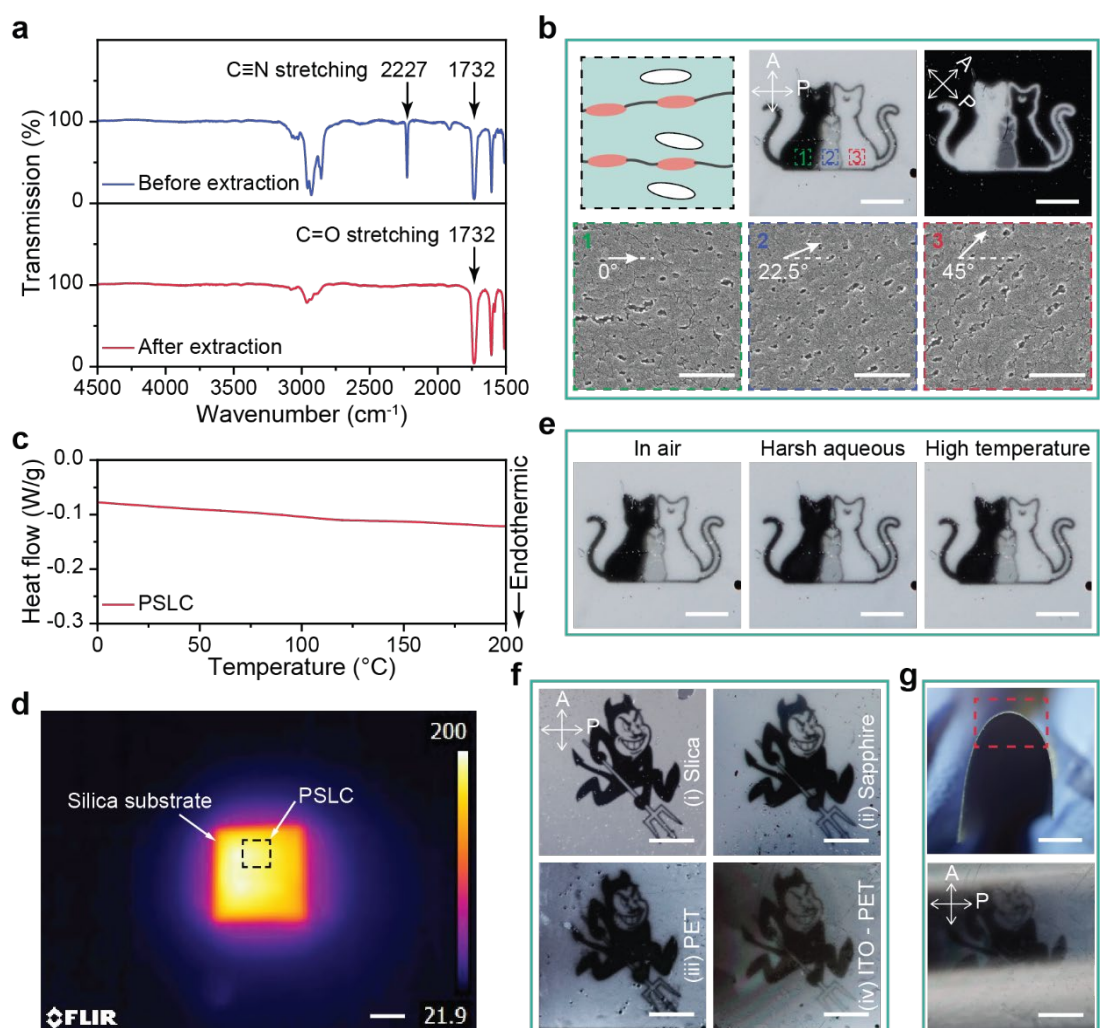

**Figure 3. Environmental robustness and broadened range-of-applicability of the photoaligned and photopatterned PSLCs.** (a) FTIR spectra of the PSLC before and after 5CB extraction. (b) Porous PSLC structure after 5CB extraction; POM images of the PSLC comprising the pattern "cat family" and corresponding SEM images of different surface regions. (c) DSC thermogram of the PSLC after 5CB extraction. (d) Thermal image of the heated PSLC on silica substrate after 5CB extraction. (e) POM images of the PSLC after 5CB extraction and subjecting to tests in harsh chemical and thermal environments. (f) POM images of the PSLCs with "sun devil" photoalignment pattern bonded on various silane-treated substrates. (g) Optical and POM images of the PSLC bonded on bent ITO-PET film. Scale bars: 10 mm (d), 2 mm (g),1 mm (all POM images) and 500 nm (all SEM images).

## 2.3. Optical Encryption and Anti-counterfeiting Enabled by Single or Cascade PSLCs

With the well-developed procedure, here we present two distinct paradigms (single and cascade) that thoroughly reveal the satisfactory capabilities of the photoaligned and photopatterned PLSCs in optical anti-counterfeiting and data encryption. Our results indicate that a single PSLC can encode complex implicit orientational and explicit geometrical information when subjected to polarized illuminations (**Figure 1**). As an example, a photopatterned explicit "ASU" logo encoded with implicit, polarized "pitchfork" pattern was fabricated and could serve as anti-counterfeiting watermark (**Figure 1a** and **Figure S12**). Detailed information, such as the QR code, can also be directly encoded onto the PSLC (**Figure 1b(i)**), whereas they are challenging to counterfeit via techniques such as mask lithography. Despite the decent performance, using single PSLC as anti-counterfeiting features is still subject to unignorable risk of information leakage, since all the optical elements that constitute the information are engraved onto one object, hence still threatened by theft or loss. To further fortify the information security, we explored the viability of information encryption using a pair of exclusive, complementary PSLCs. Firstly, we proposed a special case by adopting two orthogonal alignment directors ($\vec{n}$), $45°$ and $135°$, as an example and designed an array of square PSLC "pixels" to encode the information for the proof-of-concept demonstration. As schemed in **Figure 4a**, while being examined separately, photopatterned PLSC "pixels" with $\vec{n} = 45°$ (annotated as "$+1$" state) or $\vec{n} = 135°$ (annotated as "$-1$" state) alignment would remain optically transparent and identical with respect to each other, which was quantitatively verified by the polarized transmission spectra in **Figure 2d** and **Figure S14a**. If being placed in a cascade, spatially overlapping manner and examined (detailed schematic in **Figure S13**), a pair of two orthogonally aligned PLSCs ($45°$ and $135°$ in this case) will eliminate the transmission of linearly polarized light and the "pixel" appeared dark (annotated as "$+1 \cdot -1 = 0$" state). Oppositely, a pair of PSLCs aligned along the same director will allow the linearly polarized light to transmit through. As shown in **Figure 4a**, a pair of PSLCs both aligned along $45°$ will be optically identical to that aligned along $135°$, and the cascade "pixels" appeared bright (annotated as "$+1 \cdot +1 = +1$" or "$-1 \cdot -1 = +1$" state). The identicalness of these bright "pixels" was also quantitatively verified by the polarized transmission spectra in **Figure S14b**. We observed a color change in the bright "pixels", which indicates the spectral tuning effect from multiple birefringent layers.[67,68]

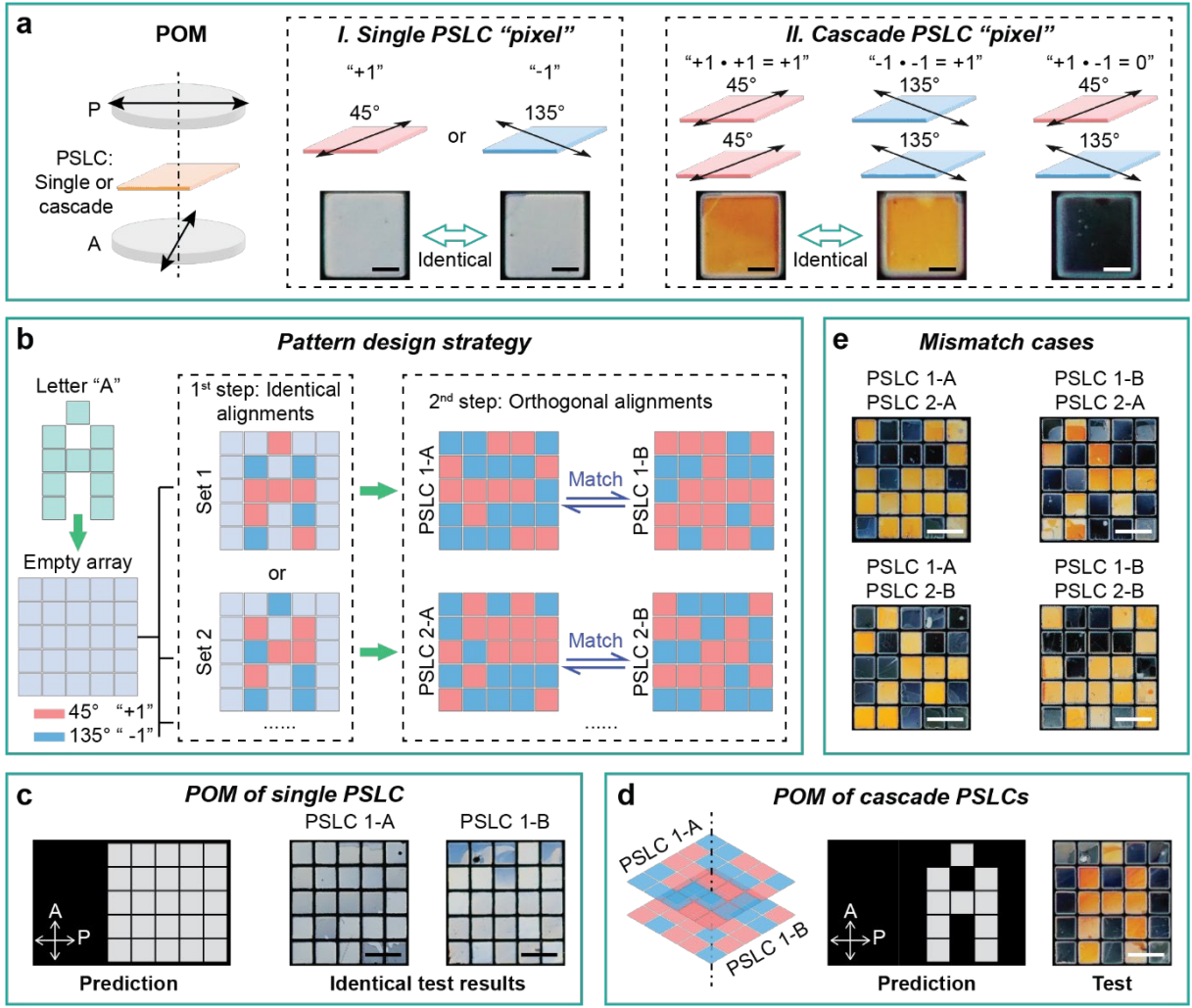

**Figure 4. Proof-of-concept optical data encryption using cascade PSLCs with 45° and/or 135° alignments.** (a) Schematic of the single or cascade PSLC "pixels" adopted for data encryption. (b) Design strategy for different sets of complementary PSLCs encoded with a letter "A". (c) Predicted and experimental results of Set 1 under separate examination. (d) Predicted and experimental results of Set 1 under cascade examination. (e) Experimental results of mismatched sets under cascade examination. Scale bars: 200 μm (a) and 1 mm (c-e).

To effectively utilize the cascade PSLC "pixels" for optical data encryption, we developed a pattern design strategy that allows us to fabricate a series of complementary PSLC sets comprising aforementioned "pixels" at "*+1*" or "*-1*" status. As depicted in **Figure 4b**, to design the complex patterns encoded with the to-be-disclosed information (*i.e.*, letter "A") for pairs of PSLCs, chosen "pixels" on the first PSLC (annotated as PSLC 1-A, PSLC 2-A, …) are first randomly filled with alignment along either *45°* or *135°*. And the spatially corresponding "pixels" on the second PSLC (annotated as PSLC 1-B, PSLC 2-B, …) are encoded with identical alignments. In the second step, each of the leftover "pixels" on both samples are randomly filled with orthogonal alignments for deceptive purposes. Experimentally, we fabricated those designed sets of complementary samples (PSLC 1-A & 1-B and PSLC 2-A & 2-B) encoded with the patterns in **Figure 4b**. In principle, both samples appear identically as an array of bright "pixels" under separate POM examinations (**Figure 4c** and **Figure S15a**) and did not disclose any meaningful information. The colored appearings on the PSLCs were defects attributed to the non-uniform sample thickness introduced during cell assembly. The localized thickness variations lead to disturbed phase separations during photopatterning, which disrupted the local alignment.[69,70] To unveil the encoded information, matched samples must be assembled in a cascade and spatially overlapping manner and examined (~500 μm spacing in between, detailed schematic in **Figure S13**), and the encoded letter "A" was disclosed, as depicted in **Figure 4d** (PSLC 1-A & 1-B) and **Figure S15b** (PSLC 2-A & 2-B). Cascade examinations on the random combinations using mismatched samples only led to meaningless patterns, which validated the exclusiveness of this design methodology (prediction in **Figure S15c** and experimental results in **Figure 4e**). Note that the variations in the color saturations of the POM images under cascade examination were caused by the sample defects and further amplified by cascade assembly.[68] Therefore, one is not able to differentiate between these two matched samples or extract meaningful information by stealing and examining only one of them, which can further prevent information leakage. Another advantage of the proposed design strategy is that it allows fabricating numerous pairs of complementary PSLCs encoded with identical information, yet are exclusive among different pairs because of the randomness, and such exclusiveness can be drastically enhanced with the increment of total number of pixels or alignment directors. Moreover, the cascade PSLCs could also retain their designed functionality after undergoing the harsh organic/aqueous environments as well as the thermal-stability tests, as confirmed by the POM images in **Figure S16**.

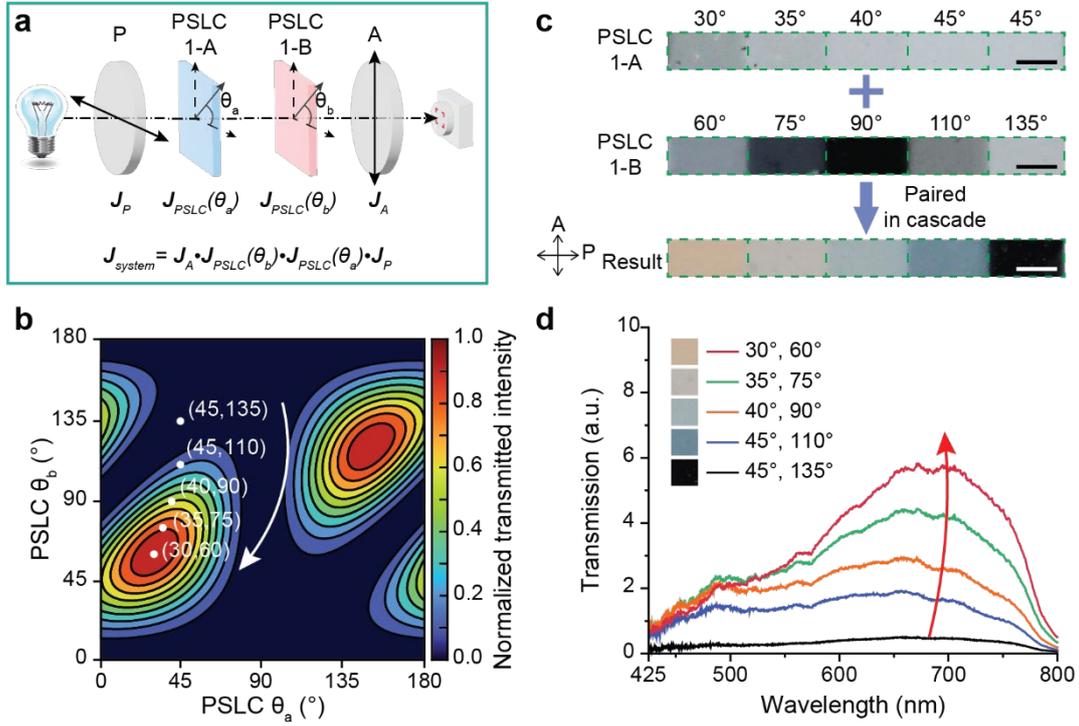

**Figure 5. Generalized design of data encryption using cascade PSLCs with programmable outputs.** (a) Diagram of optical path for calculating the outputs of the cascade PSLCs. (b) Contour of the normalized transmitted intensities using a pair of PSLCs with directors ($\vec{n}$) within the region of $0° \leq \theta_a, \theta_b \leq 180°$. (c) POM images of separately and cascade examined PSLCs with the selected combinations of $\theta_a$ and $\theta_b$, scale bar: 250 μm. (d) Polarized transmission spectra of the corresponding cascade PSLCs in (c).

To further generalize the concept of optical data encryption via paired PSLCs and deliver information with greater bit-depth, we have demonstrated the programming of the resultant output of cascade PSLC "pixels" by manipulating the directors ($\vec{n}$) of each "pixel". We adopted Jones matrix method to analytically guide our design of cascade PSLCs.[71] As illustrated in **Figure 5a**, the Jones matrix of a system consisting of a pair of cross polarizers (*0° polarizer* and *90° analyzer*, respectively) and two PSLC "pixels" (denoted as PSLC 1-A and PSLC 1-B with directors ($\vec{n}$) along $\theta_a$ and $\theta_b$, respectively) was derived to calculate the transmitted light intensity according to **Equation 1** to **Equation 9** in **Experimental Section**. Based on the calculated results, the contour of the transmitted light intensities ($I_{out}$) corresponding to arbitrary combinations of $\theta_a$ and $\theta_b$ within the region of $0° \leq \theta_a, \theta_b \leq 180°$ was plotted in **Figure 5b**. Particularly, 5 different combinations of $\theta_a$ and $\theta_b$, from (*45°, 135°*) to (*30°, 60°*) were picked as a representative set of programmable outputs. The calculated contour clearly reveals that these 5 combinations give monotonically increasing transmitted intensities. PSLC samples (annotated as PSLC 1-A and PSLC 1-B, respectively) were fabricated, each containing 5 different regions that were photoaligned accordingly. While being examined separately, POM images of PSLC 1-A and PSLC 1-B evidently show regions with distinctive brightness induced by varying LC directors ($\vec{n}$) (**Figure 5c**). While being examined in a spatially overlapping manner as a cascade, distinctive colors that emerged from the corresponding regions were observed, as shown in **Figure 5c**. The color change can be attributed to the superposition of the accumulated, wavelength-dependent phase retardations of the broadband, polarized light transmitting through multiple optically birefringent layers, and this can be verified by the corresponding polarized transmission spectra with well-resolved upward trend in the long-wavelength range in **Figure 5d**.[72–74] Therefore, we have achieved programmable spectral tuning based on cascade PSLCs. Down this path, it is expected that more output colors are readily attainable based on the systematic design of the photoaligned and photopatterned PSLCs. Hence, the optical outputs of two or even more cascade PSLCs can be efficaciously programmed by tuning the incident light, refractive indices, thicknesses as well as alignments of the PSLCs, so that a multi-colored palette could be generated based on the spectral responses. By further combining this palette with the abovementioned pixelated design methodology, one is able to encode elaborate graphical information with greater bit-depth that can fully utilize the director-dependent characteristics of PSLCs per specific designs.[75,76]

## 3. Conclusion

In this work, we presented a facile strategy to fabricate PSLCs via projection photoalignment of azobenzene dye, micro-scale alignment of LC molecules, and photopatterning of photocurable LC mixtures, aiming at achieving localized, theoretically pixel-level, and multi-directional alignment (< 50 μm) of PSLCs for optical data encryption and anti-counterfeiting purposes. Benefiting from the dynamic projection optical setup and surface silanization treatment, the PSLCs can be directly deployed onto various rigid and flexible substrates to broaden the range-of-applicability in IoTs devices. Furthermore, once extracting the embedded, non-reactive LC molecules from the PSLCs after photopatterning, enhanced chemical and thermal stability were obtained. In terms of application, these PSLCs possessing both explicit solidified geometries and implicit orientational characteristics are apt to carry and disseminate elaborate, encoded information, or function as distinctive watermarks against theft and tampering. More importantly, we for the first time developed a novel strategy of utilizing a pair of cascade PSLCs to carry the distributed and encrypted information to further fortify the security. The paired PSLCs comprising spatially programmable alignment patterns necessary to constitute the encoded information, so that the information can only be revealed when the paired PSLCs are assembled and examined. Last but not least, we analytically and experimentally showed that the cascade PSLCs could achieve programmable spectral tunings, which further endows their potential in carrying exclusive and graphical information with greater bit-depth and grants their aptitudes in their direct integrations with IoTs networks.

**Experimental Section**

*Materials*: Chemicals utilized in this work are all commercially available and were used as received. 4-Cyano-4'-pentylbiphenyl (5CB, 99%), 2-Methyl-1,4-phenylene bis(4-(3-(acryloyloxy)propoxy)benzoate) (RM257, 97%) were purchased from Aaron Chemicals. Brilliant Yellow (BY) was purchased from TCI America. 3-(Trimethoxysilyl)propyl methacrylate (TMSPMA, 98%), acetic acid (glacial, ACS reagent, ≥99.7%) 2-(2H-Benzotriazol-2-yl)-6-dodecyl-4-methylphenol (Tinuvin 171), N,N-Dimethylformamide (DMF, anhydrous, 99.8%), polyethylene terephthalate (PET) films (0.1 mm nominal thickness), indium-tin-oxide(ITO)-coated PET films (0.5 mil nominal thickness), silica cover slides, pH buffer solutions (Certified 3.00, 11.00, Certipur), and dichloromethane (DCM, ACS reagent, ≥99.5%) were purchased from Sigma – Aldrich. Ethanol (99.5%, anhydrous, 200 proof, ACROS Organics), 2-propanol (IPA, Certified ACS, Fisher Chemical), acetone (IPA, Certified ACS, Fisher Chemical), deionized water (DI water, ACS) and toluene (Certified ACS, Fisher Chemical) were purchased from Fisher – Scientific. c-plane (0001) sapphire substrates were purchased from MSE Supplies. Omnirad 2100 photoinitiator was kindly donated by IGM Resins.

*Preparation of the Photocurable Liquid Crystal (LC) Mixtures*: 5CB and RM257 were first weighed in the weight ratio of 75/25 wt%, and Omnirad 2100 and Tinuvin 171 were weighed to yield weight ratios of 1 wt% and 1.5 wt% with respect to the final mixture, respectively. The chemicals were then dissolved in excess amount of DCM via vortex mixing and underwent filtration to get rid of impurities using syringe filters (0.2 μm mesh Nylon, Fisher – Scientific). Finally, DCM was removed in a vacuum oven (Across International) at room temperature.

*Preparation of the Photoalignment Substrates*: 1.5 wt% BY was dissolved in DMF via vortex mixing and syringe filtered. Silica substrates were first cleaned with excess amount of IPA and acetone for at least two cycles and then blow-dried with clean dry air. The cleaned substrates then underwent oxygen plasma for 45 min (PE-50, Plasma Etch, Inc.). The photoalignment substrates were prepared by spin-coating BY onto the oxygen-plasma treated substrates for 35 s at 1500 RPM, after which the substrates were baked at 90 °C for 45 min and stored under vacuum prior to use.

*Preparation of the TMSPMA-Treated Substrates*: Various substrates were first cleaned, and oxygen-plasma treated similarly. For silica and sapphire substrates, ~2 mL TMSPMA was dispersed in 100 mL ethanol and sonicated for 45 min, to which 6 mL diluted acetic acid (10 vol% in DI water) was added. The substrates were then soaked in the mixture under sonication for 4 h. For bare PET and ITO-coated PET substrates, the substrates were placed in a vacuum desiccator and vapor silanized at the presence of a few droplets of TMSPMA for 1.5 h at room temperature. After functionalization, the TMSPMA-treated substrates were thoroughly cleaned with excess amount of ethanol and blow-dried with clean dry air, then stored under vacuum prior to use.

*Customized Photoalignment and Photopatterning Setup*: A light engine (Pro4710, Wintech Digital) equipped with a 405 nm UV light source and a Digital Micromirror Device (DMD, Texas Instruments) with resolution of 1920 × 1080 was used as the optical input to generate the photoalignment and photopatterning patterns. A UV lens (UV8040BK2, Universe Optics) was used to project the generated patterns onto the constructed cells. The customized optical setup yields a lateral resolution of $8.9 \times 8.9$ μm$^2$ pixel$^{-1}$ and a maximum lateral printing area of $17.09 \times 9.61$ mm$^2$. A CMOS camera (MU2003-BI, AmScope) was used to monitor the focusing status of the projected patterns. For the photoalignment, a UV linear polarizer (LPUV100-MP2, Thorlabs) carried by motorized cage rotator (K10CR1, Thorlabs) was mounted underneath the cells, whose transmission axis was controlled by the rotator. The photoalignment was conducted with linearly polarized UV light at fixed light intensity of 15.3 mW cm$^{-2}$ for 10 min per alignment region. For the photopatterning, the UV linear polarizer was removed, and the constructed PSLC cells were irradiated with unpolarized UV light at fixed light intensity of 3.4 mW cm$^{-2}$ for 15 s.

*Construction of the PSLC Cells*: The photoalignment substrates and TMSPMA-treated substrates were assembled using 5 μm-thick spacers (Nitto Denko Corporation). PSLC mixtures were then injected into the constructed cells at 90 °C via capillary force, after which the cells were left to cool down to room temperature. After photopatterning, the cells were carefully cracked open with a razor blade, and the bonded PSLCs were thoroughly rinsed in DMF to remove the residual BY, if any, soaked in excess amount of ethanol for 12 h (replaced with fresh ethanol every 4 h) to etch away 5CB, then blow-dried with clean dry air.

*Characterizations of the LC Mixture, Spin-Coated BY Substrates and PSLCs*: Differential Scanning Calorimetry (DSC) characterizations of the LC mixture as well as the PSLCs were performed in Discovery 250 (TA Instruments) with a temperature ramp of 5 °C min$^{-1}$ under nitrogen atmosphere to determine the $T_{NI}$ or to monitor any other thermal transitions. Rheological characterization of the LC mixture was performed in Discovery HR 30 (TA Instruments) using a sandblasted 20 mm parallel plate with a gap thickness of 1000 μm. Fourier-Transform Infrared Spectroscopy (FTIR) analyses on the PSLCs before and after 5CB extraction were performed on PSLCs bonded onto the silane-treated c-plane sapphire substrates. Thermal Gravimetric Analysis (TGA) was performed in TGA 5500 (TA Instruments) with a temperature ramp of 10 °C min$^{-1}$ under nitrogen atmosphere. Polarized transmission spectra of the PSLCs were obtained using a spectrometer (Andor 303iB) as well as an inverted microscope (Zeiss AXIO Observer D1) equipped with a pair of linear polarizers (LPVISE100-A, Thorlabs), schematic of the measurement setup was illustrated in **Figure S6**. In the polarized transmission measurements, either one bare substrate (corresponding to the single PSLC cases) or a cell assembled with two bare substrates (corresponding to the cascade PSLC cases) was used as the reference. SEM images of the polymerized RM257 networks were obtained by Helios 5 UX SEM/FIB (Thermo Scientific). POM images of the spin-coated BY and PSLCs were obtained by using either 1) a white light LED (Thorlabs) as the light source and captured by a digital camera (Sony A6100), or 2) an inverted microscope equipped with halogen light source (Nikon Eclipse Ti2) and a CMOS camera (MU2003-BI, AmScope), both with a pair of linear polarizers.

*Tests on the Environmental Robustness of the Photoaligned and Photopatterned PSLCs*: The photoaligned and photopatterned PSLC bonded on silica was first soaked in excess amount of ethanol for 12 h (replaced with fresh ethanol every 4 h) to etch away 5CB, leaving only polymerized RM257 behind, then vacuum dried for 4 h. The sample was then immersed in ~ 100 mL of the solvents or pH buffers, 48 h each (replaced with fresh solvents/pH buffers every 8 h), to test its environmental robustness, and was vacuum dried for 4 h in between the switching of each solvent/pH buffer. POM images of the sample were taken after being tested in each solvent/pH buffer and vacuum dried via the aforementioned setup. For the thermal stability test, the sample was heated to 200 °C using a hotplate (Fisher Scientific) and annealed there for 2 h. POM images of the sample after the thermal stability test were taken via the aforementioned setup. A thermal imaging camera (FLIR E8) was used to monitor the temperature of the sample.

*Calculations on the Angle-Dependent Transmission Output of A Pair of Cascade PSLC "Pixels"*: As depicted in **Figure 5a**, to analyze the output of a pair of cascade PLSC "pixels", the system was considered to be comprising of a pair of cross polarizers (with *0°* polarizer and *90°* analyzer, respectively) and two PLSC "pixels" (denoted as PLSC 1-A and PSLC 1-B with alignments along $\theta_a$ and $\theta_b$, respectively). Jones matrix method was used to analytically calculate the transmitted intensities.[71] Matlab coding was used for angular sweep on the results and to generate the corresponding contour.

The PSLC "pixels" were modeled as optically anisotropic components, whose Jones matrix is[8,77–79]

PSLC $\quad J_{PSLC}(\theta) = \begin{bmatrix} \cos\theta & -\sin\theta \\ \sin\theta & \cos\theta \end{bmatrix} \begin{bmatrix} X & 0 \\ 0 & Y \end{bmatrix} \begin{bmatrix} \cos\theta & \sin\theta \\ -\sin\theta & \cos\theta \end{bmatrix}$ (1)

where $\theta$ is the in-plane angle of director, $X$ and $Y$ are the complex transmission coefficients. As a proof-of-concept demonstration, here we omitted the wavelength-dependent phase retardations as the broadband light transmitting through the PSLCs. Hence, the PSLCs were assumed to be perfect, planar linear polarizing components, and $X = 1$ and $Y = 0$ were used. The Jones matrix corresponding to the entire system is

System $\quad J_{system} = J_A \cdot J_{PSLC}(\theta_b) \cdot J_{PSLC}(\theta_a) \cdot J_P$ (2)

where the Jones matrices for the corresponding components are

Polarizer $\quad J_P = \begin{bmatrix} 1 & 0 \\ 0 & 0 \end{bmatrix}$ (3)

PSLC 1-A $\quad J_{PSLC}(\theta_a) = \begin{bmatrix} \cos\theta_a & -\sin\theta_a \\ \sin\theta_a & \cos\theta_a \end{bmatrix} \begin{bmatrix} 1 & 0 \\ 0 & 0 \end{bmatrix} \begin{bmatrix} \cos\theta_a & \sin\theta_a \\ -\sin\theta_a & \cos\theta_a \end{bmatrix}$ (4)

PSLC 1-B $\quad J_{PSLC}(\theta_b) = \begin{bmatrix} \cos\theta_b & -\sin\theta_b \\ \sin\theta_b & \cos\theta_b \end{bmatrix} \begin{bmatrix} 1 & 0 \\ 0 & 0 \end{bmatrix} \begin{bmatrix} \cos\theta_b & \sin\theta_b \\ -\sin\theta_b & \cos\theta_b \end{bmatrix}$ (5)

Analyzer $\quad J_A = \begin{bmatrix} 0 & 0 \\ 0 & 1 \end{bmatrix}$ (6)

So that, for an input beam with Jones vector of

Input $\quad E_{in} = \begin{bmatrix} E_x \\ E_y \end{bmatrix}$ (7)

the output beam is

Output $\quad E_{out} = J_{system} \cdot E_{in}$ (8)

The transmitted intensity of the output is thus calculated by

Intensity $\quad I_{out} = E_{out}^T \cdot E_{out}$ (9)

where $E_{out}^T$ is the transpose matrix of $E_{out}$.

**Acknowledgments**

This work is funded by Arizona State University (ASU) startup funding, National Science Foundation (NSF) Future Manufacturing (FM) Award (CMMI 2229279), and ACS Petroleum Research Fund Award (FP00030762). The authors acknowledge the use of facilities in the Eyring Materials Center at Arizona State University. The authors gratefully appreciate Prof. Bruno Azeredo for sharing the plasma cleaner used for oxygen plasma treatment. The authors also thank Prof. Timothy Long's group in the ASU Biodesign Center for Sustainable Macromolecular Materials and Manufacturing for providing access to TGA.

**Author Information**

X.C. and K.J. conceived the idea and supervised the research, S.L., S.A., W.W., S.F., Y.Z., and L.L. conducted the experiments, S.L., X.C., and K.J. prepared the manuscript. All authors contributed to data analyses, manuscript revision and approved the submission.

Correspondence to Kailong Jin, Kailong.Jin@asu.edu; Xiangfan Chen, Xiangfan.Chen@asu.edu.

**Conflict of Interest**

The authors declare no competing interests.

**Data Availability Statement**

All data needed to evaluate the conclusions in the paper are present in the paper and/or the Supplementary Information.

**Supplementary Information**

Supplementary **Figure S1** to **S16**

    **Figure S1.** Schematic of the optical setup for photoalignment and photopatterning.

    **Figure S2.** Schematic of the photoalignment patterns, including (a) "ASU" logo encoded with ASU "pitchfork", (b) QR code, (c) ASU "Pitchfork", (d) concentric rings, (e) hierarchical chessboard, (f) cat family, (g) ASU sun devil.

    **Figure S3.** Chemical structures of the azobenzene dye and LCs.

    **Figure S4.** DSC and rheological characterizations of the photocurable 75/25 wt% LC mixture.

    **Figure S5.** (a) Transmission spectrum of photoaligned and photopatterned PSLC. Inset: Optical image of photopatterned and photoaligned PSLC with "walking cat" geometry, scale bar: 2 mm, and (b) corresponding optical profilometric profile of the photopatterned "walking cat", scale bar: 1 mm.

**Figure S6.** (a) Schematic of the optical setup for the characterization of polarized transmission spectra. (b) POM images of the uniaxially aligned PSLC, scale bar: 500 μm.

**Figure S7.** TGA result of the PSLC after 5CB extraction (i.e., neat crosslinked RM257).

**Figure S8.** Thermal stability test on the photoaligned and photopatterned PSLC. (a) photoaligned and photopatterned PSLC sample bonded onto the silane-treated silica substrate, scale bar: 5 mm. (b) POM images of the PSLC after the thermal stability test, scale bar: 1 mm.

**Figure S9.** POM images of the same PSLC sample in **Figure 3b,e** and **Figure S8** after being tested in different solvents/pH buffers, scale bar: 1 mm.

**Figure S10.** POM images of the silane-treated (a) silica, (b) c-plane sapphire, (c) PET, and (d) ITO-coated PET substrates.

**Figure S11.** Schematic illustration of the characterization of PSLCs on bent flexible substrate.

**Figure S12.** Optical profilometric profile of the photoaligned and photopatterned "ASU" watermark, scale bar: 2 mm.

**Figure S13.** Schematic illustration of the assembly and examination of cascade PSLCs.

**Figure S14.** Polarized transmission spectra of the (a) single or (b) cascade PSLC pixels with 45° and/or 135° alignments.

**Figure S15.** (a) Predicted and experimental results of Set 2 (PSLC 2-A & PSLC 2-B) under (a) separate and (b) cascade examinations. (c) Predicted results of mismatched sets under cascade examination. Scale bar: 1 mm.

**Figure S16.** (a) POM images of the complementary PSLCs encoded with letter "A" examined before and after harsh environment tests, scale bar: 1 mm.

Supplementary Information

# Robust Optical Data Encryption by Projection-Photoaligned Polymer-Stabilized-Liquid-Crystals

*Siying Liu[1,2], Saleh Alfarhan[2], Wenbo Wang[1], Shuai Feng[2], Yuxiang Zhu[1], Luyang Liu[1], Kenan Song[1], Sui Yang[2],* **Kailong Jin[2*]**, *Xiangfan Chen[1*]*

1. School of Manufacturing Systems and Networks, Arizona State University, Mesa, AZ 85212, US.

2. School for Engineering of Matter, Transport & Energy, Arizona State University, Tempe, AZ 85287, US

Corresponding Authors: Kailong Jin: Kailong.Jin@asu.edu; Xiangfan Chen: Xiangfan.Chen@asu.edu

**This PDF file includes:**

    Supplementary **Figure S1** to **S16**

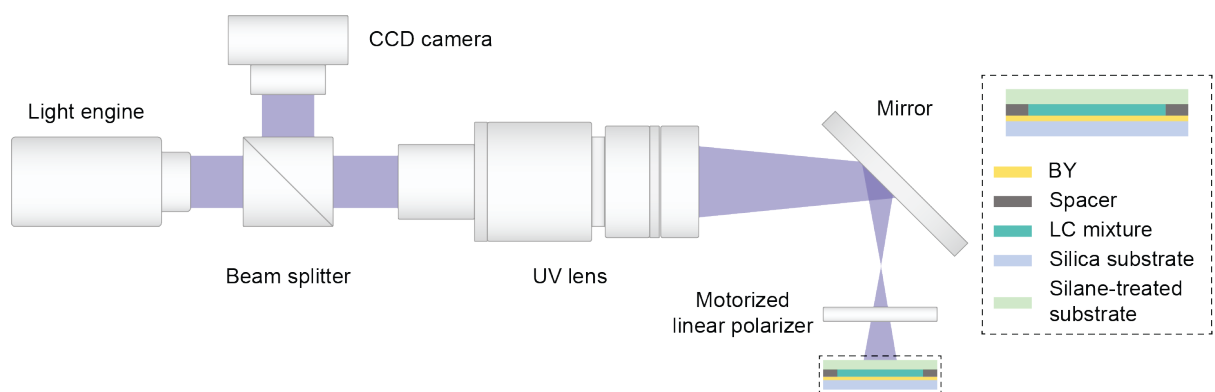

**Figure S1.** Schematic of the optical setup for photoalignment and photopatterning.

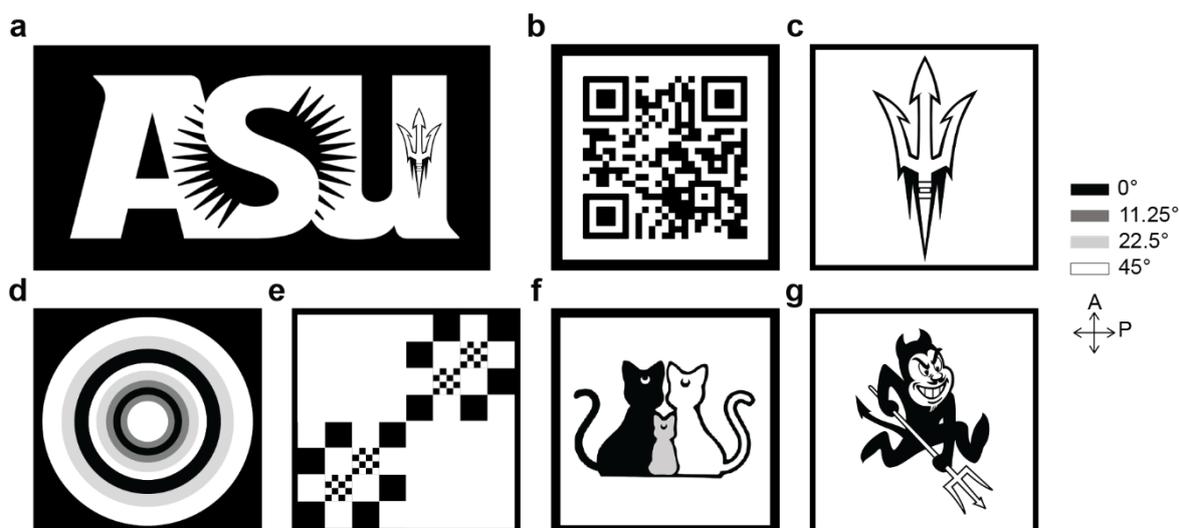

**Figure S2.** Schematic of the photoalignment patterns, including (a) "ASU" logo encoded with ASU "pitchfork", (b) QR code, (c) ASU "Pitchfork", (d) concentric rings, (e) hierarchical chessboard, (f) cat family, (g) ASU sun devil.

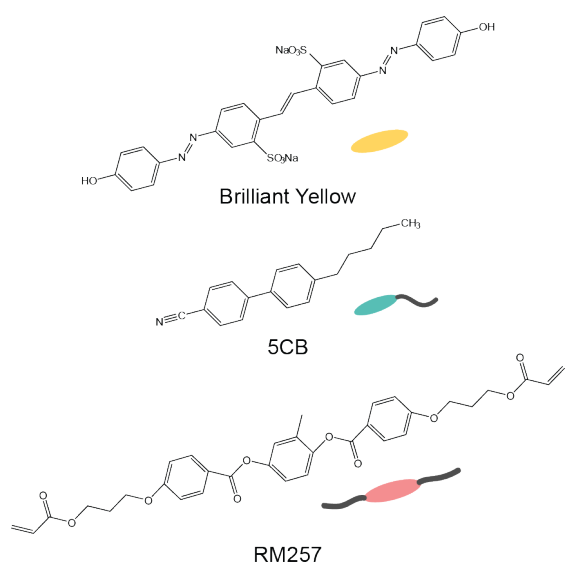

**Figure S3.** Chemical structures of the azobenzene dye and LCs.

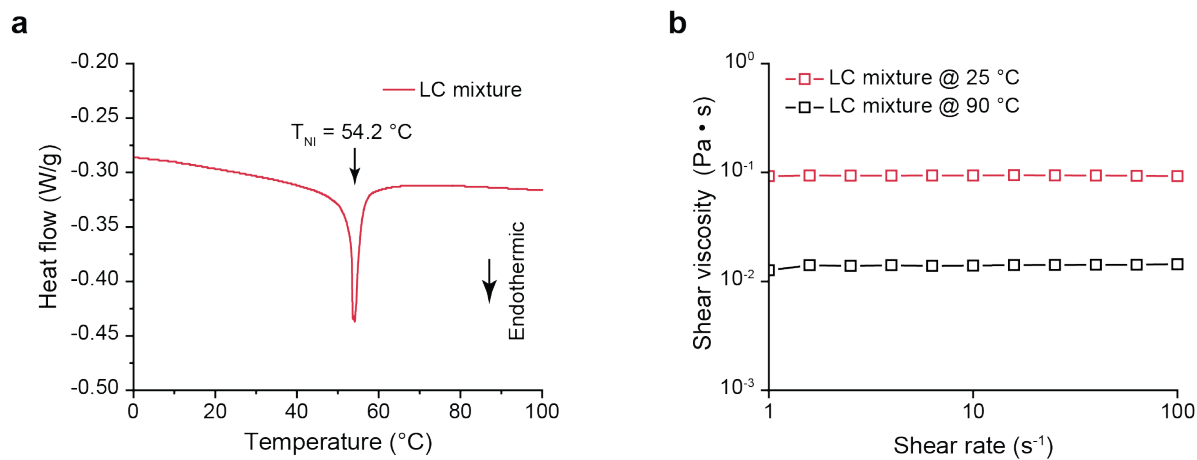

**Figure S4.** DSC and rheological characterizations of the photocurable 75/25 wt% LC mixture.

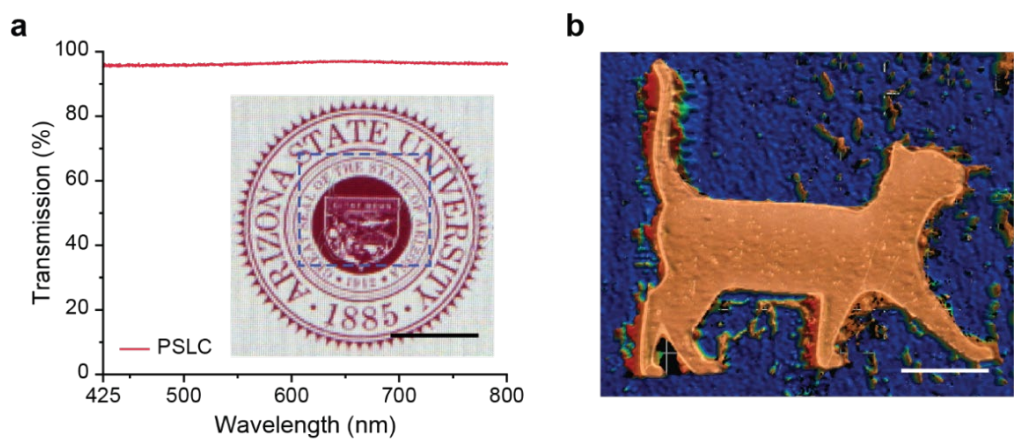

**Figure S5.** (a) Transmission spectrum of photoaligned and photopatterned PSLC. Inset: Optical image of photopatterned and photoaligned PSLC with "walking cat" geometry, scale bar: 2 mm, and (b) corresponding optical profilometric profile of the photopatterned "walking cat", scale bar: 1 mm.

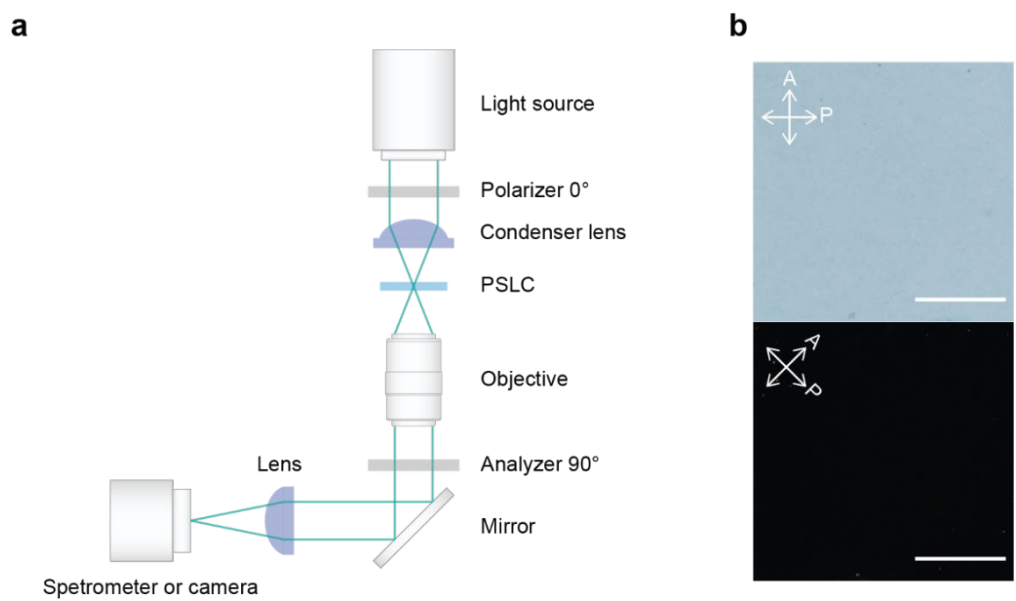

**Figure S6.** (a) Schematic of the optical setup for the characterization of polarized transmission spectra. (b) POM images of the uniaxially aligned PSLC, scale bar: 500 μm.

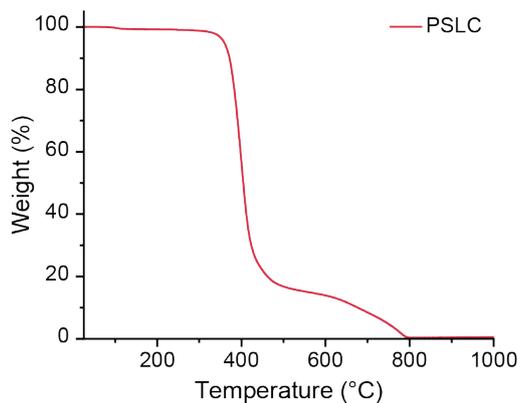

**Figure S7.** TGA result of the PSLC after 5CB extraction (i.e., neat crosslinked RM257).

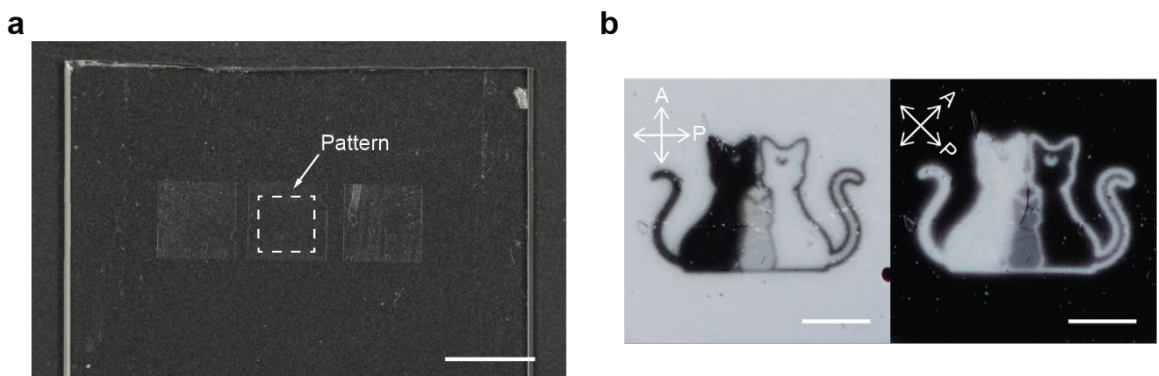

**Figure S8.** Thermal stability test on the photoaligned and photopatterned PSLC. (a) photoaligned and photopatterned PSLC sample bonded onto the silane-treated silica substrate, scale bar: 5 mm. (b) POM images of the PSLC after the thermal stability test, scale bar: 1 mm.

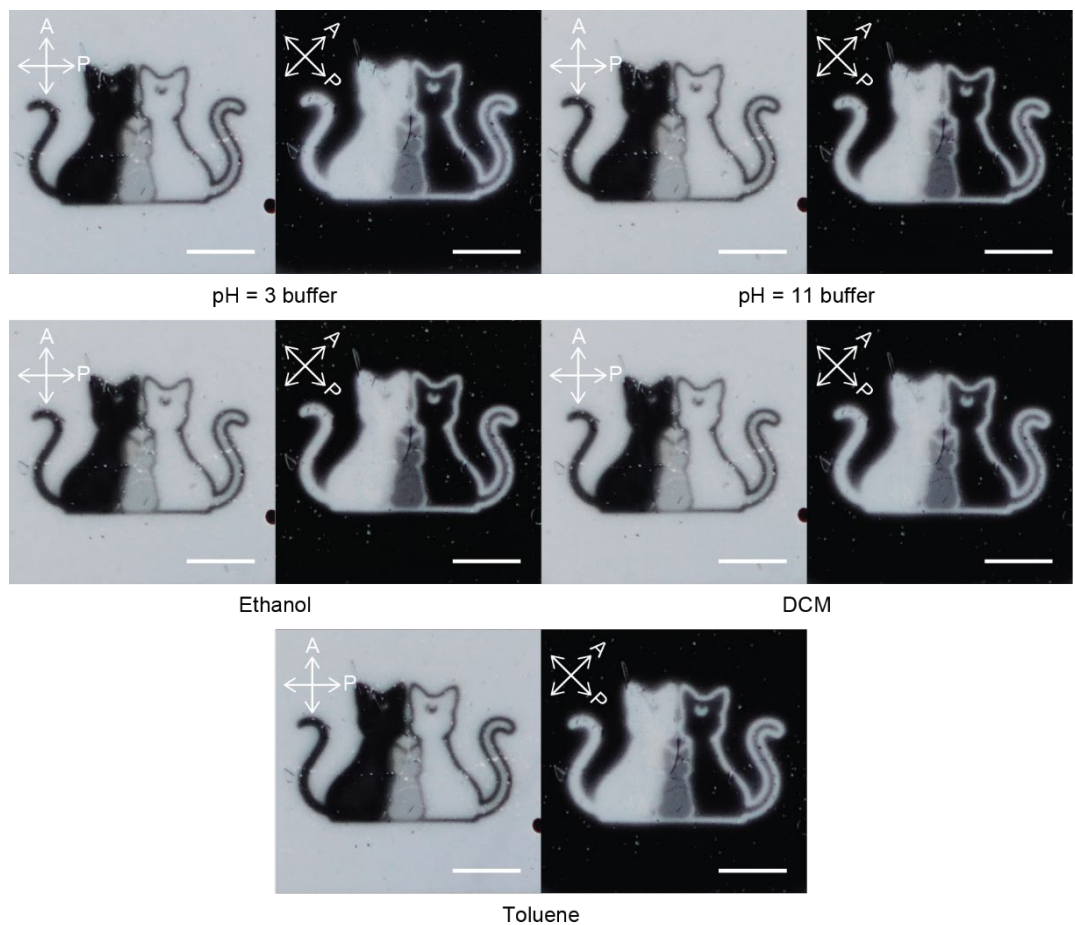

**Figure S9.** POM images of the same PSLC sample in **Figure 3b,e** and **Figure S8** after being tested in different solvents/pH buffers, scale bar: 1 mm.

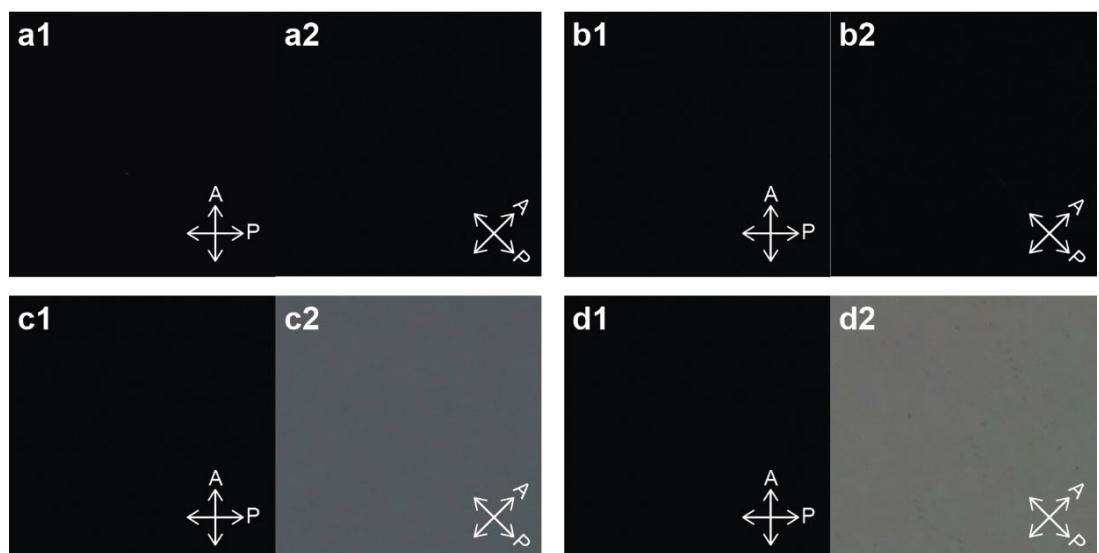

**Figure S10.** POM images of the silane-treated (a) silica, (b) c-plane sapphire, (c) PET, and (d) ITO-coated PET substrates.

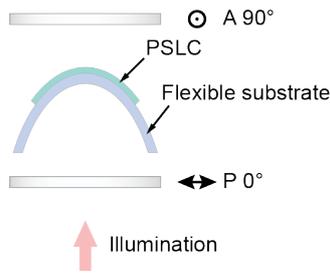

**Figure S11.** Schematic illustration of the characterization of PSLCs on bent flexible substrate.

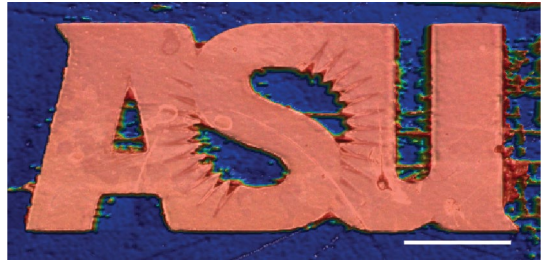

**Figure S12.** Optical profilometric profile of the photoaligned and photopatterned "ASU" watermark, scale bar: 2 mm.

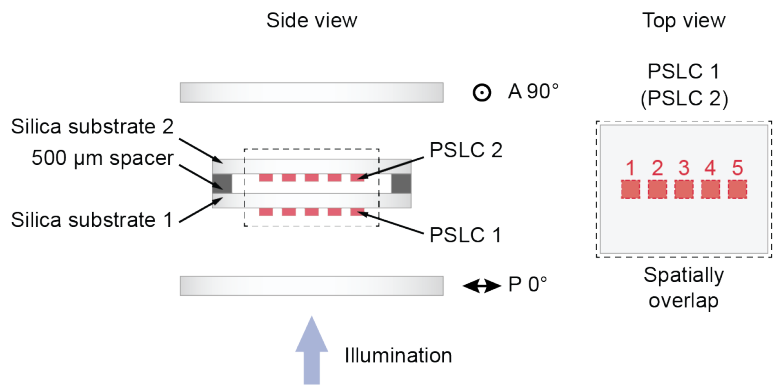

**Figure S13.** Schematic illustration of the assembly and examination of cascade PSLCs.

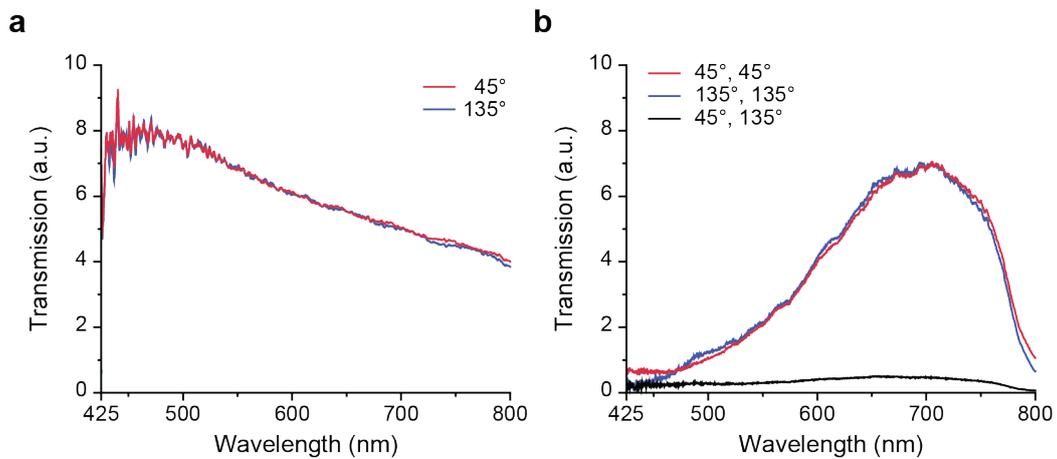

**Figure S14.** Polarized transmission spectra of the (a) single or (b) cascade PSLC "pixels" with 45° and/or 135° alignments.

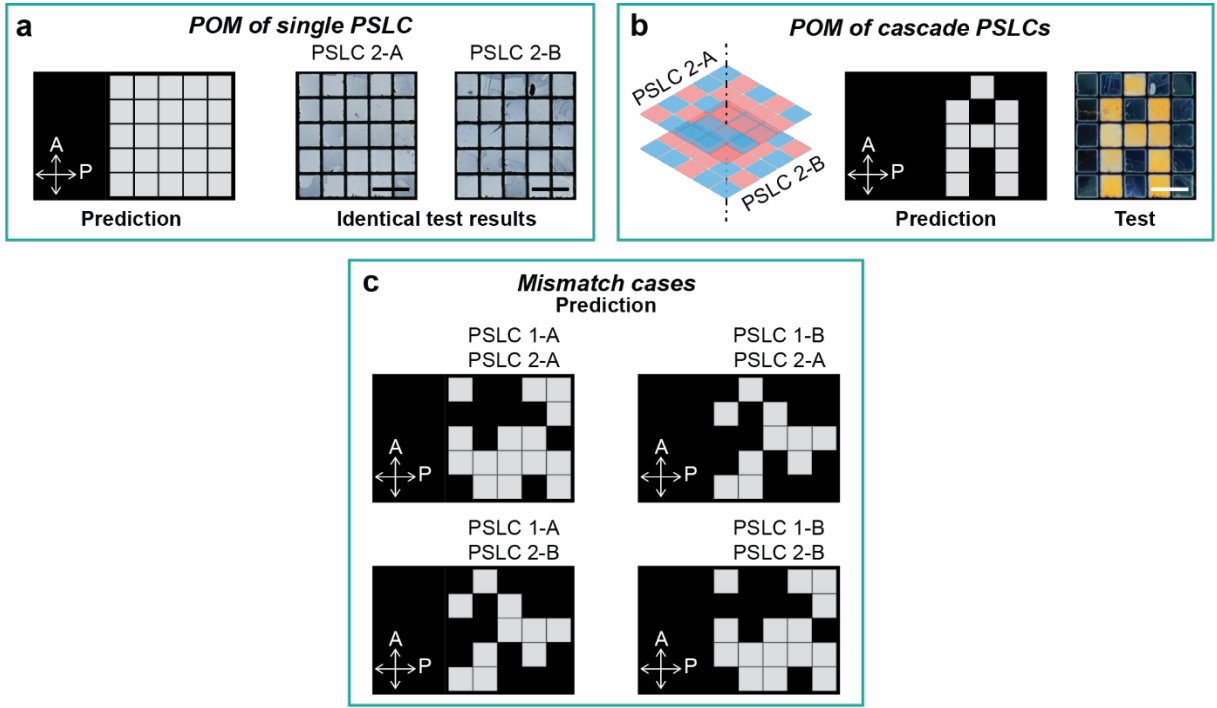

**Figure S15.** (a) Predicted and experimental results of Set 2 (PSLC 2-A & PSLC 2-B) under (a) separate and (b) cascade examinations. (c) Predicted results of mismatched sets under cascade examination. Scale bar: 1 mm.

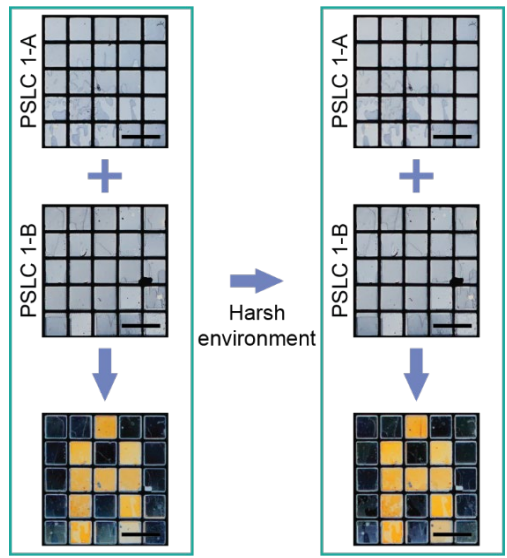

**Figure S16.** (a) POM images of the complementary PSLCs encoded with letter "A" examined before and after harsh environment tests, scale bar: 1 mm.

31